\documentclass[twocolumn,twocolappendix]{aastex631}

\usepackage{url}
\usepackage{soul}
\usepackage{multirow}

\usepackage{bm}

\begin{document}

\title{The Galactic Neutrino Sky: Predictions from Gamma-ray Source Populations} 
 
\author[0009-0004-9204-0241]{Leo W. Seen}
\affiliation{Department of Physics, Wisconsin IceCube Particle Astrophysics Center, University of Wisconsin, Madison, WI, 53706 }

\author[0000-0002-5387-8138]{Ke Fang}
\affiliation{Department of Physics, Wisconsin IceCube Particle Astrophysics Center, University of Wisconsin, Madison, WI, 53706 }

\date{\today}

\begin{abstract}

High-energy neutrino emission from the Galactic plane has been detected at a significance of $5.7\sigma$, with a prominent excess toward the inner Galaxy. We show that this excess can be naturally explained by the spatial distribution of Galactic neutrino sources. By combining $\gamma$-ray source catalogs spanning GeV–PeV energies and selecting candidate hadronic emitters, we construct ReGal-$\gamma$, a template of resolved Galactic $\gamma$-ray sources that may also produce high-energy neutrinos. Compared with models of Galactic diffuse emission, ReGal-$\gamma$ predicts a neutrino intensity that is more strongly concentrated toward the inner Galaxy. 
Combined with models of diffuse cosmic-ray emission and unresolved $\gamma$-ray sources, the template reproduces both the spectral energy distribution and the Galactic longitudinal count profile reported by IceCube without requiring additional renormalization of the emission models.
We test this result using an independent $\gamma$-ray source template constructed from different catalogs and find that our conclusion is robust against uncertainties in source modeling. Future measurements of the energy-dependent longitudinal and latitudinal neutrino distributions will provide tighter constraints on these models and help determine the spatial distribution of Galactic neutrino sources.

\end{abstract}

\section{Introduction}

High-energy neutrinos from the Milky Way have been anticipated for decades \citep{1979ApJ...228..919S}. After more than ten years of observations, IceCube has established the Galactic plane (GP) as a neutrino source \citep{IceCubeGP}.  The measured flux is broadly consistent with the multi-messenger prediction \citep{2021ApJ...919...93F}  derived from the Galactic diffuse $\gamma$-ray emission (GDGE) measured by Tibet AS$\gamma$ \citep{TibetDiff}. This suggests that the observed emission may arise from diffuse interactions of Galactic cosmic rays, discrete sources, or a combination of the two. However, the inferred neutrino luminosity of the Milky Way is much lower than the average of distant galaxies, implying that it doesn't currently host the class of sources that dominate the extragalactic neutrino background \citep{2024NatAs...8..241F}.    

The significance of the GP signal has recently increased from $4.5\sigma$ \citep{IceCubeGP} to $5.7\sigma$ in a combined analysis that uses cascades, through-going tracks, and starting tracks \citep{IceCubeGP26}. Both analyses used spatial templates primarily designed to trace Galactic diffuse neutrino emission (GDNE). However, some of the observed signal may instead originate from individual Galactic sources. \citet{IceCubeGP26} found that the GP signal is driven largely by an excess of cascade events above 5~TeV toward the inner Galaxy, within $|l|\leq20^\circ$ and $|b|\leq15^\circ$. A similar region, but more tightly restricted in latitude, with $|l|\leq30^\circ$ and $|b|\leq2^\circ$, has been studied by ANTARES and, more recently, with two years of KM3NeT/ARCA observations, and is referred to as the Galactic Ridge in these studies \citep{Albert_2023, km3netGalacticRidge}.
ANTARES reported a mild excess, while the flux inferred by scaling the IceCube all-sky GP fit to the inner Galaxy remains well below and compatible with the upper limits from KM3NeT.

GP neutrino emission can be separated into two broad components: a diffuse component and a source component. The diffuse component is produced when the cosmic-ray sea interacts with gas in the interstellar medium, whereas the source component arises from interactions occurring within or near individual cosmic-ray accelerators. Since the first evidence for Galactic neutrinos, considerable attention has been devoted to determining the relative importance of these components. By combining $\gamma$-ray source observations with measurements of the GDGE by Tibet AS$\gamma$ and LHAASO, \citet{2023ApJ...957L...6F} concluded that the total IceCube Galactic neutrino flux is likely dominated by diffuse emission from the cosmic-ray sea and unresolved hadronic sources. Related studies have also shown that the diffuse emission  and unresolved sources could contribute substantially to the neutrino flux, especially at high energies \citep{2023ApJ...956L..44V,  2024PhRvD.109d3007A, 2024ApJ...969..161G}. Nevertheless, the contribution from resolved sources may exceed that of the Galactic diffuse emission toward the inner Galaxy.

To disentangle the source and diffuse contributions, previous studies have primarily used the integrated Galactic neutrino flux. Motivated by the recently reported excess of Galactic neutrino events as a function of Galactic longitude \citep{IceCubeGP26}, we demonstrate that spatial morphology provides an additional means of separating the two components. Although the total Galactic neutrino flux may remain dominated by diffuse emission, we show that sources can dominate the pronounced peak in neutrino intensity toward the inner Galaxy.

Since the detection of a Galactic neutrino source remains elusive, we must infer the Galactic neutrino source population using other messengers. While high-energy neutrinos are only produced through hadronic interactions involving cosmic rays, high-energy $\gamma$-rays are produced by both leptonic processes involving relativistic electrons and hadronic processes. Therefore, the measured $\gamma$-ray fluxes of individual Galactic sources provide upper limits on their associated neutrino emission. Using this multi-messenger connection, we construct a spatial template of candidate Galactic neutrino sources and compare its predicted emission with the observed neutrino distribution.

\begin{figure*}  
    \centering
   \includegraphics[width=0.99\textwidth]{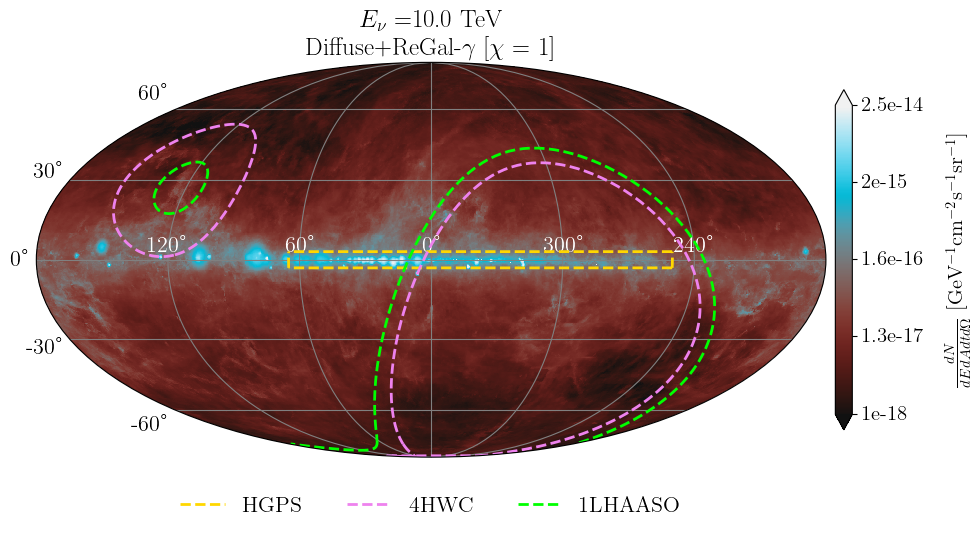}
    \caption{\label{fig:skymap}{Skymap in Galactic coordinates showing the Galactic neutrino distribution at 10 TeV obtained by adding ReGal-$\gamma$  to the CRINGE diffuse emission model, under the assumption that all $\gamma$-ray emission originates from hadronic interactions ($\chi=1$). The colorbar shows the per-flavor differential neutrino flux in units of GeV$^{-1}$cm$^{-2}$s$^{-1}$sr$^{-1}$. The regions enclosed between the two pink and green dashed lines indicate the sky coverages of the 4HWC and 1LHAASO catalogs, respectively, while the yellow dashed box marks the survey region of HGPS. ReGal-$\gamma$ also includes sources from the full-sky 4FGL-DR4 catalog.}}
\end{figure*}

We describe the spatial templates adopted for GP neutrino emission in Section~\ref{sec:TeVg}. This includes ReGal-$\gamma$, a neutrino source template constructed from resolved Galactic $\gamma$-ray sources (Section~\ref{sec:ResGalg}), an independent $\gamma$-ray source template developed for CTAO and SWGO simulations (Section~\ref{sec:SWGO}), templates of GDNE (Section~\ref{sec:GDE}) and models of unresolved sources (Section~\ref{subsec:unresolved}. Next, we convert the differential $\gamma$-ray fluxes into neutrino fluxes, predict the energy-dependent spatial distribution of Galactic neutrino emission, and compare the resulting longitudinal profiles with observations in Section~\ref{sec:nu}. Finally, we summarize our conclusions and discuss the main caveats of the analysis in Section~\ref{sec:dis}.

\section{Potential neutrino emitters based on $\gamma$-ray observations}\label{sec:TeVg}

In this section, we describe the construction of the Galactic neutrino emission model used to derive the spatial distributions of high-energy neutrinos. Our model consists of resolved $\gamma$-ray sources (Sections~\ref{sec:ResGalg} and \ref{sec:SWGO}), diffuse emission (Section~\ref{sec:GDE}), and unresolved $\gamma$-ray sources (Section~\ref{subsec:unresolved}).

\subsection{ReGal-$\gamma$ Template}\label{sec:ResGalg}

Creating a comprehensive model of resolved Galactic $\gamma$-ray sources requires detailed studies of each source's spectral and spatial distributions. As a simple approximation, we combine source catalogs covering distinct energy ranges and sky regions. Specifically, we utilize the {\it Fermi}-LAT 14-year source catalog (4FGL-DR4\footnote{DR4 specifies the fourth data release of 4FGL}) \citep{4fgl_dr4}, the H.E.S.S. Galactic Plane Survey (HGPS) \citep{HESS:2018pbp}, the First LHAASO catalog (1LHAASO) \citep{1LHAASO}, and the Fourth HAWC catalog (4HWC) \citep{4hwc} to create ReGal-$\gamma$, a GeV-PeV neutrino source model derived from Resolved Galactic $\gamma$-ray sources. 

For each catalog, we exclude extragalactic and pulsar-associated sources, including pulsars (PSRs), pulsar wind nebulae (PWNe), and TeV halos. Although some theoretical models predict a hadronic contribution from PSRs and PWNe at the highest energies, the TeV $\gamma$-ray emission from these sources is generally expected to be predominantly leptonic \citep{HAWC:2017kbo,Albert:2025gwm}. Appendix~\ref{appendix:table} summarizes each catalog and describes the corresponding source-selection process. The selected sources are then combined across different $\gamma$-ray energy ranges, as summarized in Table~\ref{tab:table2}.

An example of the ReGal-$\gamma$ template with the CRINGE diffuse emission model at $E_\nu = 10$ TeV is shown in Figure~\ref{fig:skymap}. Sources can be seen standing out from the diffuse background.

\subsection{CTA Template}\label{sec:SWGO}
A separate all-sky $\gamma$-ray source template, made for simulations of the Cherenkov Telescope Array Observatory (CTAO) and Southern Wide-field Gamma-ray Observatory (SWGO) \citep{SWGO:2025taj}, is provided by \citet{Abe_2024}. 
The template consists of sources from gamma-cat \footnote{\url{https://gamma-cat.readthedocs.io}}, the Third Catalog of Hard {\it Fermi}-LAT Sources (3FHL) \citep{3fhl}, the Second HAWC catalog (2HWC) \citep{2hwc}, and HGPS. There are 248 resolved sources and a synthetic source population (see Section~\ref{subsec:unresolved}). Spatial overlaps among the contributing catalogs were examined to remove duplicate entries. An exponential cutoff was applied to sources with TeV spectral indices harder than 2.4 to prevent unrealistic extrapolations to higher energies.

Like the ReGal-$\gamma$ template, we filter out PSRs, PWNe, and extragalactic sources. We also remove Fermi Bubbles and sources with latitudes $|b|>20^\circ$, considering that Fermi Bubbles can be well explained by inverse Compton radiation, and sources at high latitudes without classification are likely extragalactic. After filtering, the CTA template contains 100 resolved sources; the majority are SNRs and unidentified sources.  

\subsection{Galactic Diffuse Emission}\label{sec:GDE}
The GDGE has been measured across the entire sky by {\it {\it Fermi}}-LAT between 100~MeV and 1~TeV \citep{2012ApJ...750....3A, 2022ApJS..260...53A}. Above 1~TeV, the GDGE from several regions in the Northern sky has been measured by air shower $\gamma$-ray experiments, including ARGO-YBJ at $0.35-2$~TeV \citep{2015ApJ...806...20B}, Tibet AS$\gamma$ Observatory at 100-1000~TeV \citep{TibetDiff}, HAWC Observatory at $0.3-100$~TeV \citep{HAWC:2021bvb}, and LHAASO at $10-1000$~TeV \citep{2023arXiv230505372C}. These measurements can guide GDNE models, as part of the $\gamma$-ray emission should originate from hadronic interactions. 

The two GDNE models tested in the first IceCube GP analysis \citep{IceCubeGP} are the Fermi-$\pi^0$ \citep{Ackermann_2012} and KRA$_\gamma$ models \citep{Gaggero_2015}. Fermi-$\pi^0$ is a model of $\gamma$-rays from $\pi^0$ decays, derived from a cosmic-ray propagation model assuming spatially uniform cosmic-ray diffusion. The model parameters are then fit to $0.1-100$~GeV {\it{Fermi}}-LAT data. When used in neutrino analyses, the $E^{-2.7}$   $\gamma$-ray spectrum is assumed to remain constant at higher energies. Alternatively, the KRA$_\gamma$ model invokes a radially dependent cosmic-ray diffusion spectrum resulting in a harder cosmic-ray spectrum than suggested by the Fermi-$\pi^0$ model in the inner Galaxy. This yields a more peaked neutrino emission profile toward the Galactic center. 

In IceCube's most recent search for diffuse neutrinos from the GP \citep{IceCubeGP26}, the CRINGE GDNE model with an additional unresolved sources component was also tested. CRINGE is a Galactic diffuse emission model derived from solving the Galactic cosmic-ray transport equation and fitting to local cosmic-ray data \citep{CRINGE}. The resulting $\gamma$-ray profile may explain the measurements by LHAASO and TibetAS$\gamma$ in the 10-1000 TeV range when accounting for a population of unresolved sources (see Section~\ref{subsec:unresolved}). In this work, we adopt the CRINGE diffuse-emission model as our default GDNE model and discuss the impact of alternative diffuse emission models in Section~\ref{sec:dis}.

\begin{figure*}
    \centering
    \includegraphics[width=1\textwidth]{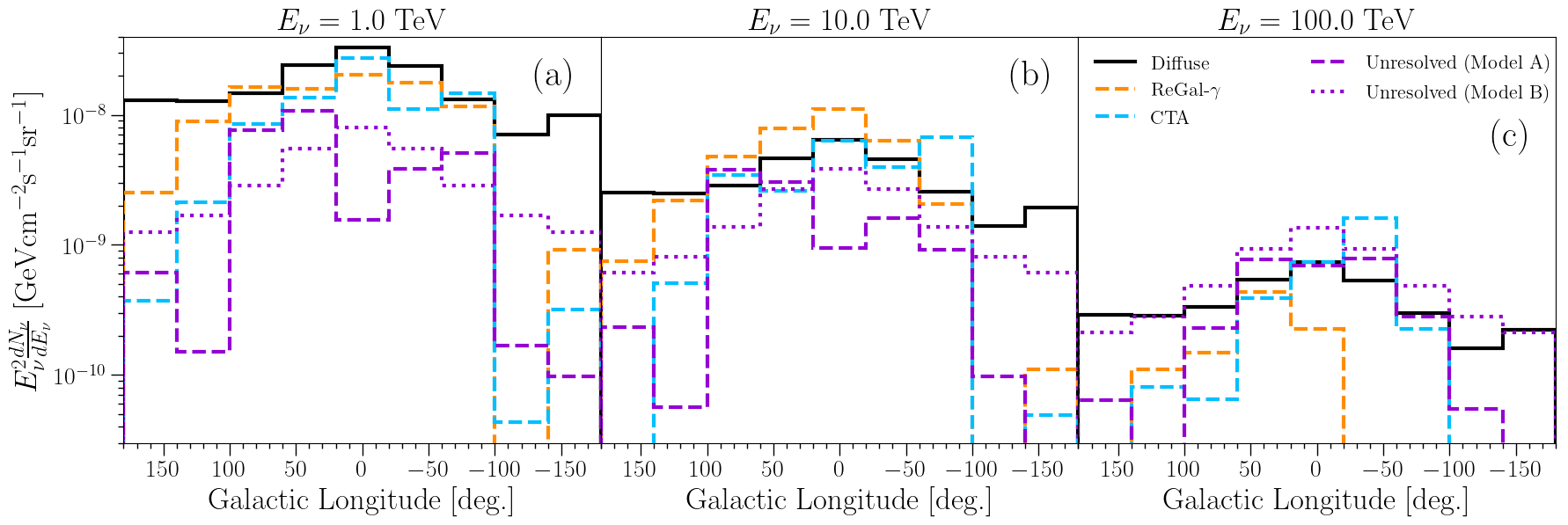}
    \caption{\label{fig:fluxProfiles}{Longitudinal profiles of the per-flavor neutrino flux from resolved $\gamma$-ray sources, represented by ReGal-$\gamma$ (orange dashed) and CTA templates (blue dashed); unresolved $\gamma$-ray sources, represented by two models (purple dashed and dotted); and diffuse neutrino emission (black solid). All source models assume $\chi=1$. Panels (a), (b), and (c) show the neutrino flux profiles at 1, 10, and 100 TeV, respectively.}}
\end{figure*}

\subsection{Unresolved $\gamma$-ray Sources}\label{subsec:unresolved}

In \citet{Abe_2024}, a population of synthetic $\gamma$-ray sources was generated by modeling the $\gamma$-ray emission from three main source classes, namely SNRs, PWNe, and $\gamma$-ray binaries, based on their birth rates and spatial distributions. The template contains 1402 synthetic sources, including  154 SNRs, 92 interacting SNRs, 1007 PWNe, and 149 $\gamma$-ray binaries. 

A fraction of the synthetic sources initially generated by \citet{Abe_2024} was removed to avoid double-counting sources from the resolved source catalogs. Since the construction of this population, 1LHAASO and 4HWC have reported additional sources. Therefore, the synthetic population provided by \citet{Abe_2024} should be regarded as a conservative upper bound on the population of sources that remain unresolved by current $\gamma$-ray telescopes. 

An alternative unresolved source model is provided by \citet{CRINGE}.  This model consists mainly of synthetic pulsars following \citet{2022ApJ...928...19V} with fluxes derived from extrapolations of luminosity and $\gamma$-ray spectra from TeV $\gamma$-ray observations. 

The unresolved source populations predicted by the two studies are broadly compatible at energies around 10~TeV. At 100~TeV, however, the unresolved flux predicted by \citet{CRINGE} is approximately a factor of two higher than that of the entire synthetic source population from \citet{Abe_2024}.

In this work, we adopt both unresolved source models to capture the associated modeling uncertainties. Model~A assumes that unresolved $\gamma$-ray sources may also emit neutrinos and uses the synthetic source population of \citet{Abe_2024}. Consistent with our treatment of resolved sources, we exclude PWNe and retain all other source classes. Model~B is based on the unresolved-source model of \citet{CRINGE}. Because sources not associated with pulsars contribute approximately 50\% of the total flux at TeV energies, we rescale the predicted flux to 50\% of its original value to remove the approximate contribution from PWN-like sources.

\section{Neutrino Emission Profiles} \label{sec:nu}

Based on the $\gamma$-ray observations in Section~\ref{sec:TeVg}, we derive the spectral and spatial distributions of high-energy neutrinos. 
For each selected source, we assume that a fraction $\chi$  of its $\gamma$-ray emission originates from hadronuclear interactions, while the remaining fraction, $1-\chi$, comes from leptonic processes. We refer to $\chi$ as the hadronic fraction. We further assume that the hadronic component has the same spectral shape as the total $\gamma$-ray spectrum. This approximation is not strictly valid, since hadronic and leptonic emission can exhibit different spectral shapes, particularly at high energies where Klein--Nishina effects become relevant. Nevertheless, it may be reasonable for the selected source classes if their emission is predominantly hadronic. 

The connection between hadronic $\gamma$-ray and neutrino emission in the Galaxy  is studied in \citet{Ahlers:2013xia,2021ApJ...919...93F,2023ApJ...957L...6F}. We derive the neutrino spectrum following  
\begin{equation} \label{eqn:gamma_nu_conversion}
    E_\nu^2\frac{dN_\nu}{dE_\nu}\bigg|_{\text{per-flavor}} \approx  \frac{\chi}{2}\left(E_\gamma^2\frac{dN_\gamma}{dE_\gamma}\bigg|_{E_\gamma = 2E_\nu}\right).
\end{equation}
Although constraints on $\chi$ have been obtained for a few sources (e.g., \citealp{Fang:2022uge, HAWC:2024kkc}), this parameter is generally difficult to determine from  $\gamma$-ray observations alone. We therefore leave $\chi$ as a free parameter. Neutrino observations provide a clean way of constraining this important quantity \citep{Fang:2024nxn}.

Figure~\ref{fig:fluxProfiles} presents the longitudinal neutrino flux distributions at 1, 10, and 100 TeV assuming $\chi = 1$ (and Figure~\ref{fig:latitudinal_flux_distribution} in Appendix~\ref{appendix:lat} presents the corresponding latitudinal distribution). In all cases, the diffuse emission dominates over the resolved sources in the outer Galaxy while the source contribution is more significant in the inner Galaxy. Overall, the ReGal-$\gamma$ template has more sources in the northern sky due to the addition of sources from 1LHAASO and 4HWC. At high energies, the ReGal-$\gamma$ template clearly reflects the scarcity of southern-sky sources resulting from the absence of air-shower observatories in the Southern Hemisphere. In contrast, the CTA template is dominated by spectral extrapolations from lower energies and is therefore subject to large uncertainties.

\begin{figure}
    \centering
    \includegraphics[width=1\columnwidth]{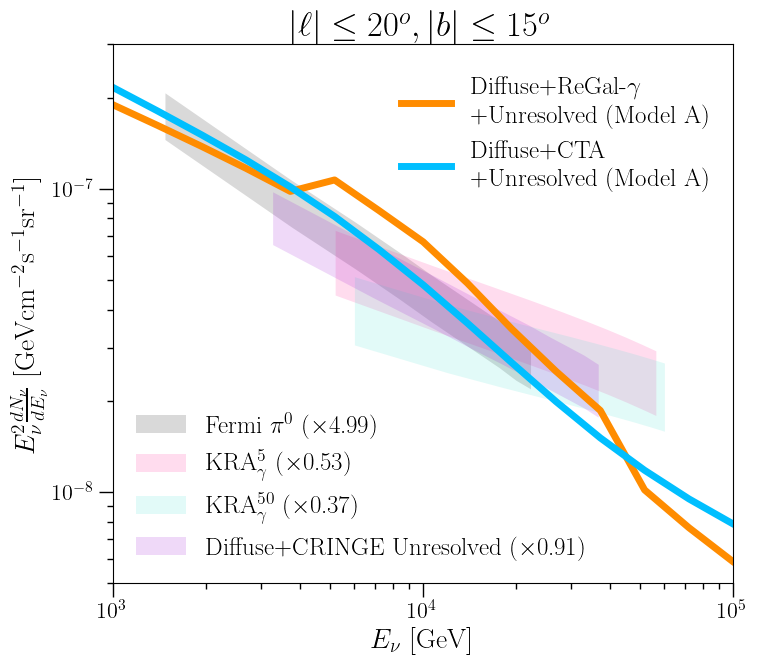}\caption{\label{fig:SEDinnerGalaxy}Average per-flavor neutrino flux in the inner Galaxy. The colored bands show the average per-flavor neutrino flux reported by IceCube for various GDNE models as calculated from an all-sky fit. The corresponding flux renormalization factors are listed in the legend. The solid orange and blue lines are the model predictions that account for resolved and unresolved sources, in addition to the CRINGE GDNE model. Both the resolved and unresolved source models assume $\chi = 1$.}
\end{figure}

In Figure~\ref{fig:SEDinnerGalaxy}, we show the neutrino energy spectrum from the inner Galaxy. The bands represent the post-fit spectra obtained for four diffuse-emission templates with different renormalizations, as reported by \citet{IceCubeGP26}. We find that the calculated inner-Galaxy flux can be explained by the sum of an unrenormalized diffuse component and the contribution from hadronic $\gamma$-ray sources, including both resolved and unresolved populations, for an average hadronic fraction of $\chi \sim 0.5$--$1$. We also show the all-sky spectral energy distributions (SEDs) of these flux templates in Figure~\ref{fig:allsky_spectrum} of Appendix~\ref{appendix:sed}. The kinks in the ReGal-$\gamma$ SED are due to the discrete energy ranges selected during template construction.

\begin{figure} 
    \centering
   \includegraphics[width=1\columnwidth]{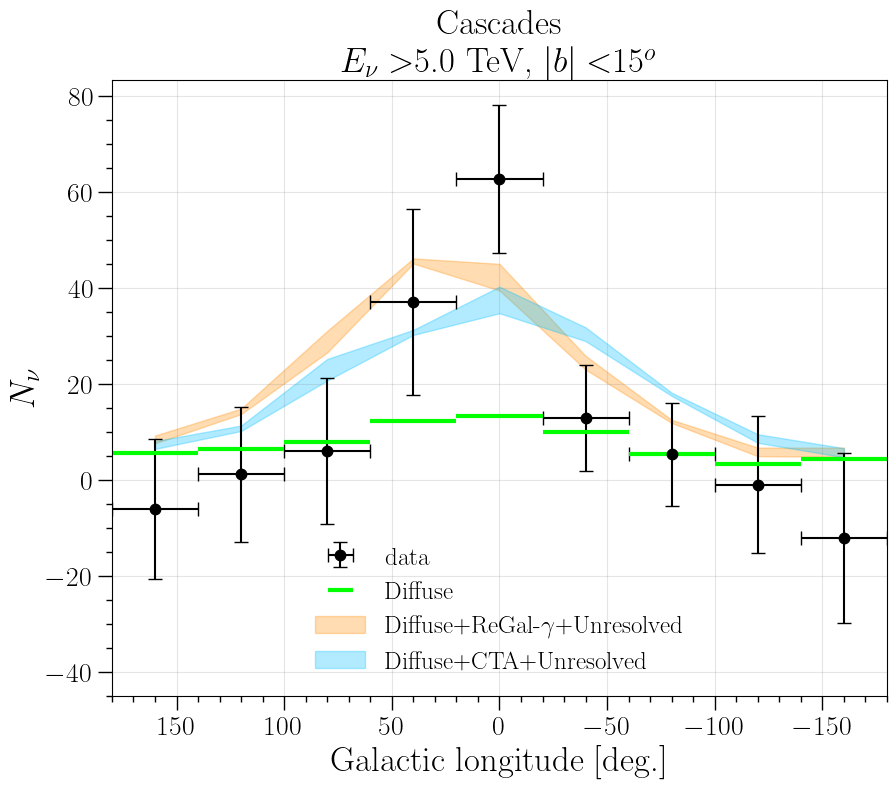}\caption{\label{fig:countsProfile}{The black data points with errorbars are observed data in nine longitudinal bins as reported in \citet{IceCubeGP26}. The green lines indicate the CRINGE diffuse model expectations. The orange and light blue shaded regions are the model predictions that account for resolved and unresolved sources, in addition to the CRINGE GDNE model. The minimum and maximum of the shaded regions indicate the uncertainties in unresolved source models.
    }}
\end{figure}

From the flux maps, we can derive the expected counts by convolving the various neutrino templates with IceCube's effective area \citep{IceCubeGP26}, $A_{\rm eff}$, which is a function of declination $\delta$ and neutrino energy, 
\begin{equation}\label{eqn:nu_counts}
    N_\nu (l, \Delta l)= \int \frac{dN_\nu}{dE_\nu dAdtd\Omega} (l, b) A_{\rm eff} (E_\nu, \delta) t_{\rm live} dE_\nu d\Omega.
\end{equation}
Additionally, to account for reconstruction effects, we smear the flux templates by 17.5$^o$ across all energies when calculating $N_\nu$ for cascades. The smearing angle is based on the angular uncertainties of the cascade events \citep{seen2025enhancingsearchesastrophysicalneutrino} and chosen to reproduce the model predictions reported by IceCube. No smearing is applied to the flux templates in the track calculation as the angular resolution of tracks is less than one degree. The longitudinal per-flavor neutrino count distribution for tracks is shown in Appendix~\ref{appendix:l_plot_track_distribution}, Figure~\ref{fig:countsProfileTrack}. We perform the integral in Equation~\ref{eqn:nu_counts} as a discrete sum over the energy-dependent templates in each sky direction. No interpolation between energy templates is performed. For the effective area, we consider two sky regions, the northern ($\delta>5^o$) and southern sky ($\delta\leq5^o$), as well as two event morphologies, tracks and cascades. We scale $N_\nu$ to the full 12-year livetime of the ICEMAN sample \citep{thiesmeyer2025measuringastrophysicalgalacticplane}, $t_{\rm live}$. 

Figure~\ref{fig:countsProfile} presents $N_\nu$ for cascade events in corresponding sky regions in comparison to observations. Both ReGal-$\gamma$ and the CTA template, when combined with the diffuse background and unresolved source populations, can explain the excess observed in the cascade sample in the inner Galaxy region where the IceCube GP fit is most sensitive. We note that most southern-sky sources are drawn from HGPS, which surveys only $|b|\leq 3^\circ$, substantially narrower than the $|b|\leq 15^\circ$ region considered in the IceCube analysis. Bright, undetected sources may therefore lie just outside the GP and could further contribute to the neutrino flux in the central bin.

\section{Discussion and Conclusions}\label{sec:dis}

We have shown that the spectral and longitudinal distributions of high-energy neutrino emission from the Galactic plane can be simultaneously explained by a Galactic neutrino emission model based on $\gamma$-ray observations. In particular, both distributions can be reproduced by combining an unrenormalized diffuse emission component with a population of $\gamma$-ray sources not powered by pulsars. The source distribution inferred from $\gamma$-ray observations is more concentrated toward the inner Galaxy compared to the diffuse emission predicted by CRINGE. This concentration may explain why the observed neutrino profile is similarly peaked in this region and why fits based solely on diffuse emission templates such as the Fermi-$\pi^0$ and KRA$_\gamma$ models require additional renormalization. 

One method to identify the relative contributions of the two components, diffuse and sources, is to simultaneously fit both the source and diffuse normalizations. This can be done for individual sources or a population of sources, such as the ones that make up the ReGal-$\gamma$ template presented here. The inclusion of a model-dependent GDNE background in neutrino analyses is analogous to how the GDGE is treated in $\gamma$-ray experiments such as HAWC and Fermi. Furthermore, additional years of data from IceCube and KM3NeT will better measure the longitudinal and latitudinal profiles and further constrain our models.

While both ReGal-$\gamma$ and the CTA template, as discussed in Section~\ref{sec:ResGalg} and Section~\ref{sec:SWGO}, aim to parameterize the spectral and spatial distributions of resolved $\gamma$-ray sources, there are still numerous differences between the templates. First, the CTA template includes older source catalogs compared to those used in ReGal-$\gamma$. Next, some of the SEDs of HGPS sources are different between ReGal-$\gamma$ and the CTA template. One example of this is for HESS~J1641-463. The energy spectrum used in the CTA template is from \citet{Abramowski_2014}, while ReGal-$\gamma$ uses the energy spectrum provided in HGPS \citep{HESS:2018pbp}. A direct comparison of the two spectra can be found in \citet{Mares_2021}. Finally, the method by which catalogs are combined and source spectra are characterized is different. Extrapolating the SED of GeV sources to TeV energies inevitably leads to overestimations of the source flux. To account for this, the CTA template implements exponential cutoffs for hard-spectrum sources, while ReGal-$\gamma$ only uses source catalogs in discrete energy ranges. Both of these methods are approximations, and to precisely model the SED across a wide energy range requires detailed per-source studies. Importantly, despite differences among the templates, the conclusion that including source populations provides a better explanation of the neutrino spatial distribution remains robust.

We caution that although the combined coverage of HGPS, 1LHAASO, and 4HWC spans almost the entire Galactic plane, the $\gamma$-ray source templates are constructed from catalogs based on different observing facilities and strategies. Consequently, the sensitivities vary across sky regions, both within individual catalogs and between catalogs. The HGPS point-source sensitivity reaches approximately $0.5$--$2.5\%$ Crab at 1~TeV. For 1LHAASO, the point-source sensitivity is approximately $1$--$10\%$ Crab for WCDA at 3~TeV, assuming a spectrum of $dN/dE\propto E^{-2.5}$, and approximately $3$--$40\%$ Crab for KM2A at 50~TeV, assuming a point-source morphology and an $E^{-3.5}$ spectrum. The sensitivity of 4HWC at 2~TeV is comparable to that of 1LHAASO. These nonuniform sensitivities inevitably introduce incompleteness and selection biases into the compiled catalog. Nevertheless, a comparison of sources observed by H.E.S.S. and HAWC \citep{2021ApJ...917....6A}, accounting for differences in point-spread functions and background-estimation methods, concluded that the $\gamma$-ray skies observed by the two instruments are consistent. Southern-sky air-shower observatories, such as SWGO, will be essential for constructing a full-sky very-high-energy $\gamma$-ray catalog with more uniform sensitivity and analysis methods. Future measurements of the energy-dependent neutrino morphology will further constrain source models derived from these $\gamma$-ray observations.

\begin{acknowledgments}
K.F. acknowledges support from the National Science Foundation (PHY-2238916, PHY-2514194) and the Sloan Research Fellowship. This work was supported by a grant from the Simons Foundation (00001470, KF). 
\end{acknowledgments}

\bibliography{references}


\appendix

 \section{Summary of the construction of ReGal-$\gamma$}\label{appendix:table}

\begin{deluxetable*}{lccc}
\centering
\tablecaption{Summary of how ReGal-$\gamma$ was constructed. The listed sky regions indicate where each source catalog is applied. This will not necessarily match the fields of view of each detector. For example, in the second energy bin, HGPS only goes up to $l<55^\circ$ instead of $l<65^\circ$ since, after filtering, the last source in this area is at $l\approx 55^\circ$. Additionally, not all longitudes within the provided range have sources. The parentheses next to the 1LHAASO catalog indicate which detector was used. The $^*$ indicates sources in 4HWC that have no 1LHAASO counterpart.   
\label{tab:table2} 
}
\tablecolumns{5}
\tablewidth{0pt}
\tablehead{\colhead{Energy Range [TeV]} & \colhead{Catalog}	& \colhead{Sky regions} } 
\startdata
$E_\gamma\leq 1$ & 4FGL  &    All-sky     \\
\hline 
\multirow{3}{*}{$1<E_\gamma\leq 10$}    & HGPS  &    $l<55^\circ$, $l>260^\circ$     \\
& 1LHAASO (WCDA) & $55^\circ<l<260^\circ$\\ 
& 4HWC$^*$& $55^\circ<l<260^\circ$\\ 
\hline
\multirow{4}{*}{$10<E_\gamma\leq 25$}    & 1LHAASO (WCDA)  & $10^\circ<l<190^\circ$ \\
& HGPS & $190^\circ<l<358^\circ$\\
& 4HWC & $l<10^\circ$, $l>358^\circ$ \\
& 4HWC$^*$ & $10^\circ<l<190^\circ$ \\
\hline
\multirow{2}{*}{$25<E_\gamma\leq 100$}    & 1LHAASO (KM2A)  &    $l<180^\circ$, $l>355^\circ$     \\
& 4HWC$^*$ & $l<180^\circ$, $l>355^\circ$ \\
\hline
$E_\gamma\geq 100$    & 1LHAASO (KM2A)  &    $10^\circ<l<180^\circ$     \\
\hline
\enddata
\end{deluxetable*}

Below, we elaborate on the catalogs used in the construction of ReGal-$\gamma$ as well as the spatial and spectral information of each source. A table summary of the ReGal-$\gamma$ construction is provided in Table~\ref{tab:table2}.

{\it 4FGL-DR4} contains 7194 $\gamma$-ray sources observed by {\it Fermi}-LAT, most of which are extragalactic. Each source entry contains both spectral and spatial information. The energy spectrum is modeled either by a power law, a log parabola, or a power law with a super-exponential cutoff, while the spatial distribution is characterized either by a point source, disk, Gaussian, or custom template model, the latter of which allows for more complex models such as ring-type distributions. Due to the large number of reported source types, we only list the selected source classes, which include the following: globular clusters ('glc'), supernova remanants ('SNR', 'snr'), star forming regions ('SFR', 'sfr'), high mass binaries ('HMB', 'hmb'), low mass binaries ('LMB', 'lmb'), binaries ('Bin', 'bin'), Galactic center ('GC'), nova ('NOV'), and special case sources ('SPP', 'spp') \footnote{Special case sources indicate potential associations with SNR or PWN. Uppercase source classes denote identified sources while lowercase source classes denote associations.}.  

{\it HGPS} contains 78 sources detected by H.E.S.S. in the inner Galaxy between Galactic longitudes $l>250$ and $l<65$ and latitudes $|b|\leq 3^\circ$. At these energies and sky regions, most sources are likely Galactic, as high-energy $\gamma$-rays from more distant sources are absorbed and attenuated. The energy spectrum of HGPS sources is modeled either by a power law or a power law with an exponential cutoff, while the spatial distribution is characterized either as a point source, Gaussian, shell, or custom template model. We remove sources labeled PWN and also HESS J1943+213, which is likely extragalactic \citep{HESS_J1943+213}. 

{\it 1LHAASO} contains 90 sources, 43 of which are detected at ultrahigh energy (UHE; $\geq100$ TeV), making 1LHAASO the largest catalog of UHE $\gamma$-ray sources to date. Similar to HGPS, the majority of these sources are Galactic. All sources in the catalog have extensions less than 2 degrees, a single power-law energy spectrum, and a Gaussian or point-like spatial profile. We exclude 1LHAASO J1219+2915, 1LHAASO J1104+3810, 1LHAASO J1653+3943, 1LHAASO J1727+5016, and 1LHAASO J2346+5138, all of which already have definitive associations with extragalactic sources or are located at high Galactic latitudes with no known Galactic counterparts. We also remove the 35 pulsar-associated sources listed in Table 4 of \citet{1LHAASO}. Furthermore, we remove 1LHAASO J0703+1405, which is spatially coincident with Monogem.

Finally, {\it 4HWC} reports a total of 85 sources, 22 of which have no 1LHAASO counterparts. Like 1LHAASO and HGPS, these sources are mostly Galactic. The energy spectrum is modeled either by a power law or a log parabola, while the spatial morphology is described either by a point source or a Gaussian template. In addition to sources with Pulsar labels, we remove 4HWC J0347-1405, 4HWC J1104+3810, 4HWC J0856+2910, 4HWC J1230+1223, and 4HWC J1654+3944, all of which have a known extragalactic association or are far off the Galactic plane with no known Galactic counterpart.

The spectral and spatial information for each source in all four catalogs except 4HWC is obtained through the Gammapy software package \citep{gammapy}. Source information from 4HWC is obtained directly from the paper. We use the best-fit spatial and spectral information reported in the catalogs. For 1LHAASO, we use the measurements from WCDA and KM2A within their respective energy ranges. 

We combine the sources from the above-mentioned catalogs in different energy ranges as follows. 

{$\bm E_\gamma\leq\bm 1 \textbf{ TeV}$:} Since ground-based $\gamma$-ray observatories have a limited field of view, we use 4FGL-DR4 up to its maximum sensitive energy range. After filtering, 262 sources remain, the majority of which lie along the GP. Most are special case sources or globular clusters. Sources with high Galactic latitudes are mainly globular clusters.

{$\bm 1\textbf{ TeV}<\bm E_\gamma\leq\bm10 \textbf{ TeV}$:} The sensitivity of H.E.S.S. peaks in this energy range. Therefore, we
adopt HGPS and fill the outer Galaxy with sources from WCDA. Additionally, we add 4HWC sources with no WCDA counterpart in the same spatial region as WCDA. After filtering, there are 88 sources: 65 from HGPS, 19 from WCDA, and 4 from 4HWC. The majority of the HGPS sources are unidentified. 

{$\bm 10\textbf{ TeV}<\bm E_\gamma\leq\bm25\textbf{ TeV}$:} In this energy range, air shower $\gamma$-ray observatories start to offer better sensitivity than IACTs. We switch to using WCDA's entire field of view while filling the rest of the sky with 4HWC and HGPS. Once again, non-overlapping 4HWC sources with WCDA are added in the WCDA region. After filtering, there are 81 sources: 35 from WCDA, 32 from HGPS, and 14 from 4HWC. 

{$\bm 25\textbf{ TeV}<\bm E_\gamma\leq\bm100\textbf{ TeV}$:} Here, we use KM2A and sources in 4HWC that are not in 1LHAASO. After filtering, there are 56 sources: 42 from KM2A and 14 from 4HWC. This is the first energy bin where declinations below $-26^\circ$  become undetectable due to the lack of air shower $\gamma$-ray detectors in the southern sky. 

{$\bm E_\gamma>\bm100\textbf{ TeV}$:} KM2A is the only catalog used in this energy bin. After filtering, 41 sources remain. As in the previous energy bin, coverage of the southern sky is limited because 1LHAASO includes only sources with declinations above $-20^\circ$.

{\bf Cygnus Cocoon:} The Cygnus X region is one of the largest nearby star-forming complexes. Both HAWC and LHAASO have detected extended $\gamma$-ray emission from this region \citep{2021NatAs.tmp...50A, LHAASOCOLLABORATION2024449}. The Cygnus Cocoon is the residual $\gamma$-ray emission obtained after subtracting two smaller sources associated with the PWN TeV~J2032+4130 and the SNR $\gamma$ Cygni. Its morphology is similar to that observed in infrared and GeV $\gamma$-ray emission \citep{Abeysekara:2021yum}. Because the Cygnus Cocoon is absent from the relevant source catalogs \footnote{Although 4HWC~J2030+4056 is spatially coincident with the Cygnus Cocoon, it is associated with 1LHAASO~J2031+4052u, which KM2A detects as a point source at the position of Cygnus~X-3. We therefore exclude 4HWC~J2030+4056 from our template to avoid double counting the corresponding 1LHAASO source.}, with the exception of 4FGL-DR4,  we include it separately in our analysis over the 1--100~TeV energy range using the spatial and spectral models reported by HAWC \citep{Abeysekara:2021yum}.

\section{Spectral energy distributions} \label{appendix:sed}
\begin{figure} 
    \centering
   \includegraphics[width=1\columnwidth]{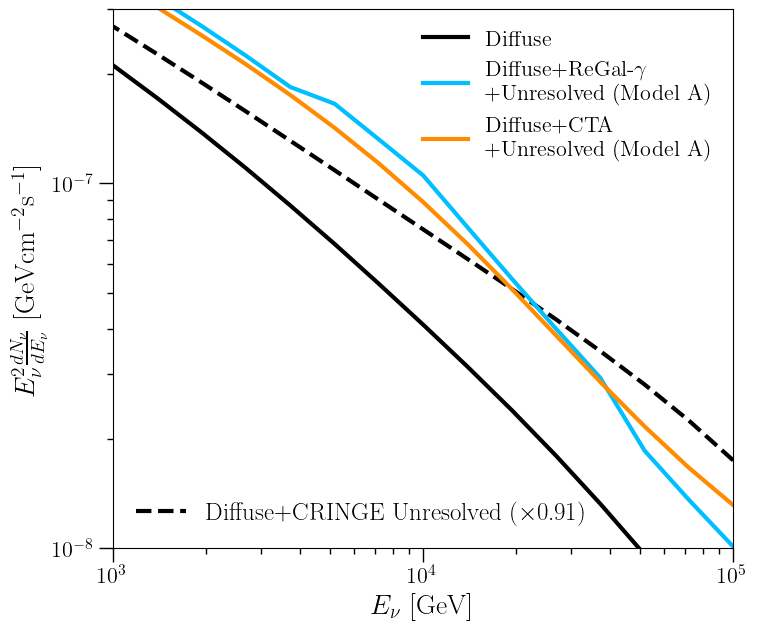}
    \caption{\label{fig:allsky_spectrum}{All-sky per-flavor neutrino flux spectrum. The black dashed line shows the energy spectrum of the combined CRINGE diffuse and unresolved-source models, normalized to the IceCube best-fit measurement. The solid lines show predictions from the diffuse emission only based on the CRINGE model (black) and diffuse emission plus resolved and unresolved sources models (orange and blue). All resolved and unresolved source models assume $\chi = 1$.}}
\end{figure}

The all-sky per-flavor neutrino spectrum is shown in Figure \ref{fig:allsky_spectrum}. For the model including ReGal-$\gamma$, while the kinks in the energy spectrum are smaller than in Figure~\ref{fig:SEDinnerGalaxy} due to the all-sky average, the remaining kinks are once again due to the discrete energy ranges chosen during template construction. Nevertheless, both the CRINGE plus resolved and unresolved sources models agree well between 1-10 TeV. To match the measured normalization by IceCube, $\chi$ can be tuned for either the resolved or unresolved sources components. 

\section{Longitudinal Tracks Distribution}\label{appendix:l_plot_track_distribution}
\begin{figure} 
    \centering
   \includegraphics[width=1\columnwidth]{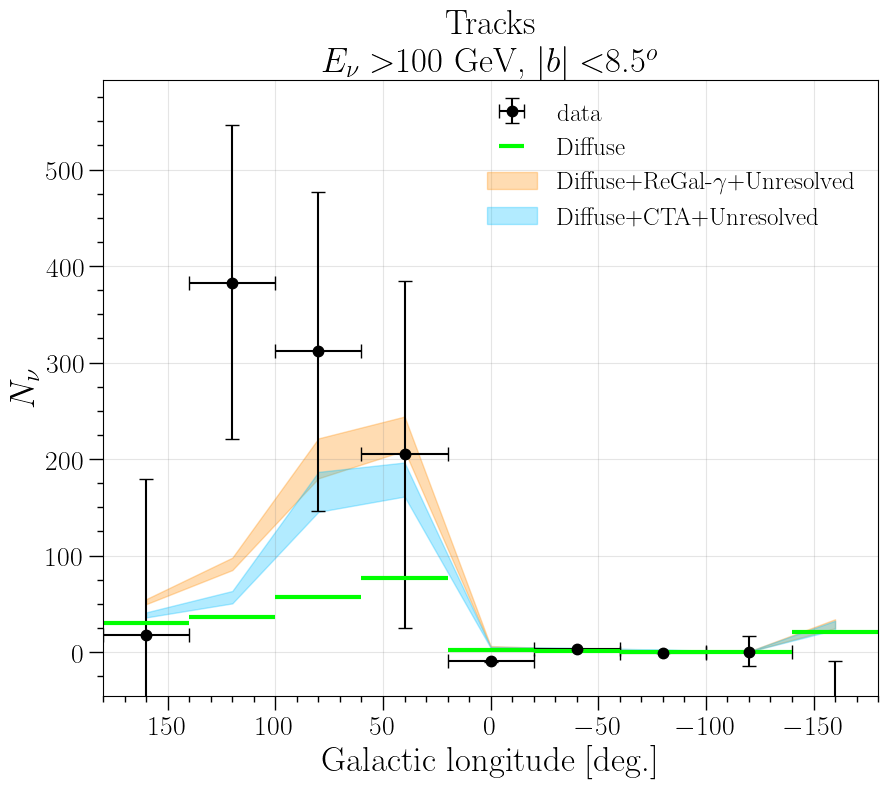}
    \caption{\label{fig:countsProfileTrack}{Same as Figure~\ref{fig:countsProfile} but for tracks. The energy cut is at lower energies and the Galactic latitude cut is tighter due to the much smaller angular resolution of tracks.}}
\end{figure}

While cascades provide information on larger extended sources, tracks provide a better probe of individual point-like sources. Carrying out the same integral as in cascades, we obtain the longitudinal counts distribution as shown in Figure~\ref{fig:countsProfileTrack}. A large excess can be seen in the $100^\circ<l<140^\circ$ bin for tracks. While the error bars for the track bins are large, it is unknown if there are additional neutrino sources that are not $\gamma$-ray bright, contributing to the observed data.

\section{Latitudinal Distributions}\label{appendix:lat}
\begin{figure*}
    \centering
    \includegraphics[width=1\textwidth]{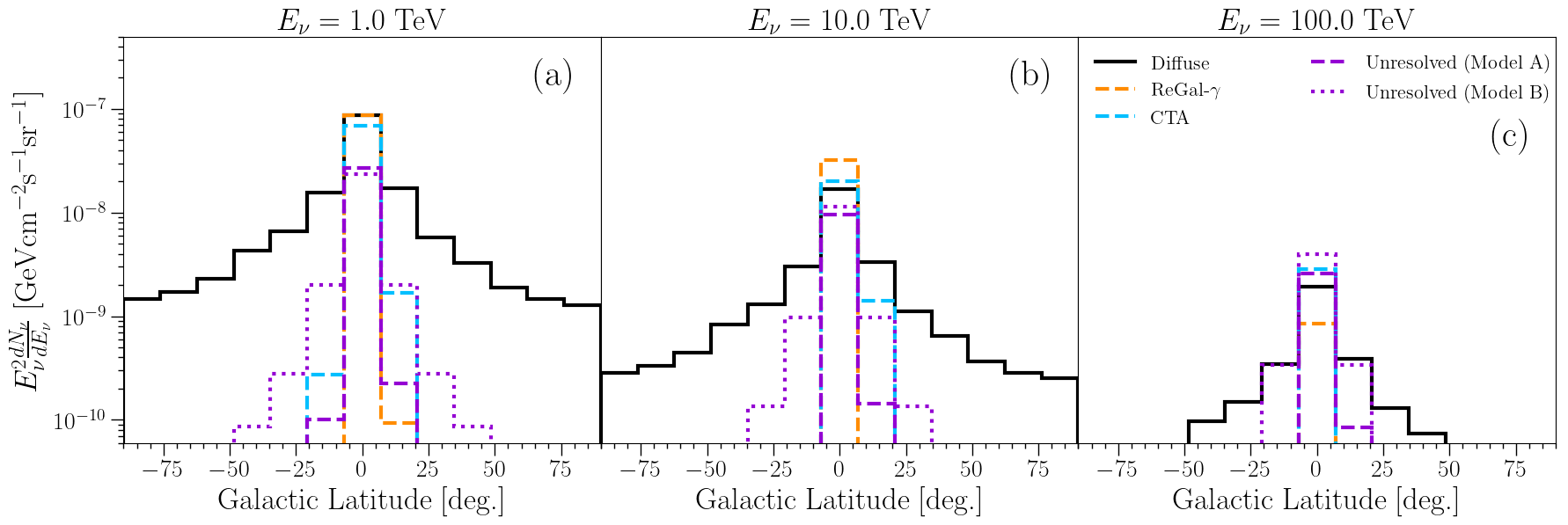}
    \caption{\label{fig:latitudinal_flux_distribution}Latitudinal per-flavor neutrino flux distribution. The solid black line shows the CRINGE diffuse model while the colored dash/dotted lines show the contribution from sources. Panels (a), (b), and (c) show the flux distributions at 1, 10, and 100 TeV, respectively.}
\end{figure*}

The latitudinal neutrino flux distributions are shown in Figure~\ref{fig:latitudinal_flux_distribution}. 
Comparing to the longitudinal distribution, the latitudinal profile is dominated by the diffuse component except in the inner tens of degrees where the GP is present. At all energies, the source flux is comparable or higher than the diffuse emission in that region. 

\end{document}